\documentclass[11pt,preprint]{aastex}                     % onecolumn (standard format)
\usepackage{graphicx}
\usepackage{natbib}
\usepackage{epstopdf}

 \usepackage{aps-bibstyle}  % use this style if you don't use BibTeX.
\begin{document}

\title{Solar Dynamics, Rotation, Convection and Overshoot%\thanks{Grants or other notes
}

%\titlerunning{Solar Dynamics}        % if too long for running head

\author{S. Hanasoge$^{1}$,
        M. S. Miesch$^2$,
        M. Roth$^3$,
        J. Schou$^4$,
        M. Sch\"{u}ssler$^4$,
        M. J. Thompson$^2$
}


\affil{
$^1$Tata Institute of Fundamental Research, Mumbai 400-005, India;
$^2$High Altitude Observatory, National Center for Atmospheric Research,
Boulder, CO 80301, USA; 
$^3$Kiepenheuer-Institut f{\"u}r Sonnenphysik,
D-79104 Freiburg, Germany; 
$^4$Max-Planck-Institut f{\"u}r Sonnensystemforschung,
Justus-von-Liebig-Weg~3,
37077 G{\"o}ttingen, Germany
}


%\maketitle

\begin{abstract}
We discuss recent observational, theoretical and modeling progress made in 
understanding the Sun's internal dynamics, including its rotation, 
meridional flow, convection
and overshoot. Over the past few decades, substantial theoretical and observational
effort has gone into appreciating these aspects of solar dynamics.
A review of these observations, related helioseismic methodology and inference and computational
results in relation to these problems is undertaken here. 
%\bfnote[Add some more specific topic details.]
\keywords{Sun \and dynamics \and rotation \and convection \and overshoot}
\end{abstract}

\section{Introduction}
\label{sec:1}
Rotation and convection play an important role in solar and stellar evolution. They also play a crucial role in generating the magnetic activity exhibited by the Sun and other stars. 

Rotationally induced instabilities, convection and overshoot mix chemical elements within stars, with possible consequences both for the observed chemical abundances at the surface of the star and for the availability of fuel for the nuclear energy generation in the star's core. Overshoot too can affect the chemical balance of the star, in the case of the Sun for example transporting a fraction of the fragile elements lithium and beryllium to regions where the temperature is sufficiently high for these elements to be destroyed. 

Rotation and convection, and their interplay, can create dynamo action and generate magnetic field in a star. In the Sun, the overshoot layer at the base of the solar convective envelope may also play a crucial role in storing the magnetic field until it is strong enough to become buoyantly unstable and rise through the convection zone to the surface. The convection interacts with the rising magnetic field and largely determines how the magnetic field appears at the surface.

Observational data on the rotation, convection and other internal dynamics of the Sun were formerly restricted to what could be observed at the surface: granulation and supergranulation set up by the convective motions, and the surface rotation rate inferred from spectroscopic measurements and from observing the motion of tracers of the rotation such as sunspots and other surface features. This situation has been revolutionized by helioseismology, the observation and analysis of global oscillations and more localized acoustic wave propagation. At the same time as helioseismology has revolutionized the observational study of the Sun's internal dynamics, advances in numerical simulations -- greatly assisted by the massive increase in computational power of modern supercomputers -- have enabled great advances in the modeling and consequent understanding of the internal dynamics of the Sun and other stars. This chapter will discuss these observational and theoretical advances.

\section{Flows in the solar interior}
\label{sec:2}

The global-scale flows in the Sun are its rotation and meridional circulation. Flows on smaller scales include convection, from granular scales to the putative giant-cell convection, and outflows and inflows around and beneath active regions. This section gives a brief overview of these different flows. The succeeding sections will then focus on what has been learned observationally about these flows from helioseismology, and the theoretical advances being made in modeling them.

\subsection{Rotation}

Like other stars, the Sun acquired its angular momentum from the interstellar gas cloud from which it formed. As the gas contracts to form a star, unless there is some extremely efficient mechanism for losing angular momentum, the proto-star spins up so as to rotate much faster than the parent cloud. Thus young stars are typically observed to be fast rotators, and that was presumably the case for the Sun also. Over the next 4-5 billion years, the Sun would have lost angular momentum from its surface layers via the solar wind \citep[see, e.g.,][for models]{bouvier13,gallet13}. If there were no 
angular-momentum transport mechanisms at play in the solar interior, this would result in a fast-rotating core and a slowly rotating envelope. 
The evidence from helioseismology (Section~\ref{sec:3}) does not support 
such a picture, indicating that the radiative interior is rotating nearly uniformly at a rate intermediate between the polar and equatorial rate of the convection zone.  This indicates that one or more mechanisms - magnetic fields, transport by rotationally induced instabilities and by gravitational waves \citep[see review by, e.g.][]{mathis13} -- have systematically extracted angular momentum from the 
radiative interior and redistributed the residual, suppressing rotational gradients.  The solar surface has long been observed to rotate differentially, with the mid- and high-latitude regions rotating more slowly than the 
equatorial region. That this differential rotation largely persists with
depth throughout the solar convective envelope is now established observationally by helioseismology. Angular momentum transport must maintain this differential rotation, and the mechanisms for that are addressed below in 
Section\ \ref{sec:mean-flows}.

\subsection{Meridional circulation}\label{sec:mc}

There is a poleward meridional circulation in the near-surface layers of the Sun in both the north and south hemispheres. Mass conservation dictates that their 
must be a return, equatorward, flow somewhere in the solar interior, and this is often presumed to occur near the base of the convection zone. However, whether
a single meridional cell spans the whole depth of the convection zone or whether it is shallower, with perhaps other meridional cells stacked beneath the one 
nearest the surface, is a matter of active discussion. There is also uncertainty about 
whether the poleward flow extends right up to the poles, or whether there might
be a counter cell at high latitudes. There are marginal observations at best to say whether or not there is any meridional circulation in the Sun's radiative interior, though there are theoretical arguments for the existence of a slow, Eddington-Sweet circulation \citep[e.g.][]{Zahn1992}.

Determining the meridional flow as a function of radius and latitude
is of considerable interest. This is to a large extent driven by
the importance of the meridional flow in certain dynamo theories.

\subsection{Convection and convective overshoot}

Energy transport in the optically thin solar photospheric layers transitions from being effected by convection to free-streaming radiation.
A spatio-temporal power spectrum of photospheric flows reveals granular and supergranular scales. Observed properties of granules, such as spatial scales, radiative intensity and spectral-line formation are highly accurately reproduced by numerical simulations \citep[e.g.,][]{stein00,voegler2005,nordlund09}.
   One may conjecture that the success of the simulations in spite of being in an entirely different parameter 
regime is due to the small scale height ($\sim$200 km; compare with radius of the Sun, $R_\odot \sim 700, 000$ km), 
leading to the strong expansion of convective upflows, thereby smoothing 
out large fluctuations and making them almost laminar \citep{Nordlund1997}. The effects of turbulence are restricted to the downflows, where
they appear to not have a big impact on the observable near-surface dynamics. Thus incorporating the ingredients of an 
accurate equation of state and background stratification lead to a high-fidelity reproduction of line formation and spatial scales. 

Ostensibly, modeling convection in the solar interior presents a more formidable challenge, since there appear to be no 
dominant physical ingredients that are concurrently computationally tractable.
Interior convection is likely governed by aspects more difficult to model, such as the integrity of descending plumes to diffusion and
various instabilities \citep{rast98}. Further, solar convection is governed by extreme parameters \citep[Prandtl number $\sim 10^{-6} - 10^{-4}$, Rayleigh number $\sim 10^{19} - 10^{24}$, and Reynolds number $\sim 10^{12} - 10^{16}$;][]{miesch05}, which makes
fully resolved three-dimensional direct numerical simulations impossible for the foreseeable future. It is likewise difficult to reproduce 
solar parameter regimes in laboratory experiments.

%%%%%%%%%%%%%%%%%%%%%%%%%%%%%%%%%%%%%%%%%%%%%%%%%%%%%%%%%%%%%%%%%%%%%%%%
Phenomenology such as mixing-length theory (MLT) treats convective transport
as being effected by parcels of fluid of specified spatial and velocity scales, coherent over a length scale 
termed the {\it mixing length}. 
Despite its overt simplicity \citep{cox_2004}, it has been remarkably successful as an integral component of stellar structure
models \citep[e.g.][]{DiMauro10} and in describing the sub-photospheric stratification and heat transport in local solar convection simulations \citep{Trampedach:Stein:2012}.  Because 
density and pressure scale heights increase with depth, MLT posits a corresponding increase in spatial convective scale (while velocities reduce), suggesting the existence of large convective cells, the so-called {\it giant cells}.  
Three-dimensional simulations of global convection have been performed at increasing resolutions over the past few years \citep{miesch_etal_08,charbonneau10,kapyla1, kapyla2,Guerrero13b,Hotta14a,Hotta14b}, most invoking the {\it anelastic approximation} \citep{gough69}, a regime describing low-Mach number strongly non-Boussinesq stratified convection.  These simulations predict large-scale convective turnover and velocity amplitudes comparable to those of MLT. 
 %The convective velocities seen in these simulations are comparable (in a narrow sense) to MLT predictions.
Considerable effort has been spent in attempting 
surface \citep{hathaway00,hathaway13} and interior detection \citep{duvall, duvall03} of giant cells 
but unambiguous identification remain elusive.  This difficulty in detection suggests that the amplitudes
of giant cells may be significantly less than predicted by MLT and convection simulations, potentially
posing serious challenges to our understanding of deep solar convection (section 3.3).

At the base of the convection zone, convective motions are quickly decelerated by the
steep subadiabatic stratification of the radiative interior.  This thin overshoot 
region (section 3.3) coincides with the rotational shear layer known as the solar 
tachocline (section 3.1) as first posited by \cite{Spiegel1992},
possibly suggesting a causal connection.  The dynamics in this region 
are complex, involving penetrative convection and internal waves interacting with 
a stably-stratified, magnetized shear flow and the much longer dynamical time scales 
of the radiative zone \citep{hughe07}.

%%%%%%%%%%%%%%%%%%%%%%%%%%%%%%%%%%%%%%%%%%%%%%%%%%%%%%%%%%%%%%%%%%%%%%%%

\section{Helioseismic constraints on the Sun's internal dynamics}
\label{sec:3}

One of the great successes of helioseismology has been to measure the 
rotation of much of the solar interior, excluding the solar core and a region around the Sun's rotation axis. Much of this has been achieved using 
global-mode helioseismology \citep[e.g.,][]{Deubner1984}, which proceeds 
by the analysis of the observed properties of global resonant modes of 
oscillation of the Sun, in particular the mode frequencies.
See for example the excellent review by \cite{Howe2009}.
Global-mode helioseismology has been complemented by the development and application of a number of other approaches known collectively as 
local helioseismology. These include time-distance helioseismology
\citep{duvall}, 
acoustic holography
\citep{Lindsey1997}, 
and ring-diagram analysis
\citep{Hill1988}. 
Local helioseismology has been particularly used to probe the Sun's meridional (i.e. north-south) circulation
and convective motions in the upper part of the Sun's convection zone, and motions around and under active regions \citep[e.g.,][]{Komm2005}. 

\subsection{Rotation }

The global-mode frequencies $\omega_{nlm}$ of the Sun are labeled by the radial order $n$, and the degree $l$ and azimuthal order $m$ of the spherical harmonic that describes the horizontal structure of the mode. In the absence of rotation and other departures from perfect spherical symmetry, the frequencies would be independent of $m$. The Sun's rotation lifts that degeneracy. The difference in frequency between modes with like values of $n$ and $l$ but different values of $m$ is called the splitting. For a star with internal rotation rate $\Omega (r, \theta)$ (with respect to spherical polar coordinates $r$, $\theta$, $\phi$), 
the rotational splitting is given by \citep{Hansen1977}
\begin{equation}
\delta\omega_{nlm} \equiv \omega_{nlm} - \omega_{nl0}
= 
-i\int\rho\left(\xi_{nlm}^\ast\cdot{\partial\over\partial\phi}\xi_{nlm}\right)
\Omega\, d V
\Big{/}
\int
\rho\xi_{nlm}^\ast\cdot\xi_{nlm} d V
\end{equation}
where $\xi_{n;m}$ is the radial displacement eigenvector of the mode, $\rho$ is the density, and the integrals are over the interior volume of the star. This can be conveniently rewritten as
\begin{equation}
\delta\omega_{nlm} 
= 
m \int_{0}^R \int_{0}^\pi
K_{nlm}(r,\theta) \Omega(r,\theta) r d\theta dr
\end{equation}
where the $K_{nlm}$ are known as rotational splitting kernels. If the static structure of the stellar interior is known, the kernels can be calculated, 
so that the only unknown in equation (2) is the rotation rate $\Omega(r,\theta)$. Measurements of the Sun's mode frequencies therefore provide a set of observational constraints of the form (2), which may be used to infer (or at least constrain) the solar internal rotation rate.

The solar internal rotation rate is thus fairly well determined from the photosphere down to beneath the convection zone, excluding a region around the poles. 
See e.g. \cite{Thompson1996}, \cite{Schou1998}, \cite{Howe2011}, \citet{korzennik13} for examples of results using 
data from, respectively, GONG \citep{Harvey1996}, 
MDI \citep{Scherrer1995}
and HMI \citep{hmi}.

\begin{figure}[!ht]
\centering
\includegraphics*[width=8.0cm]{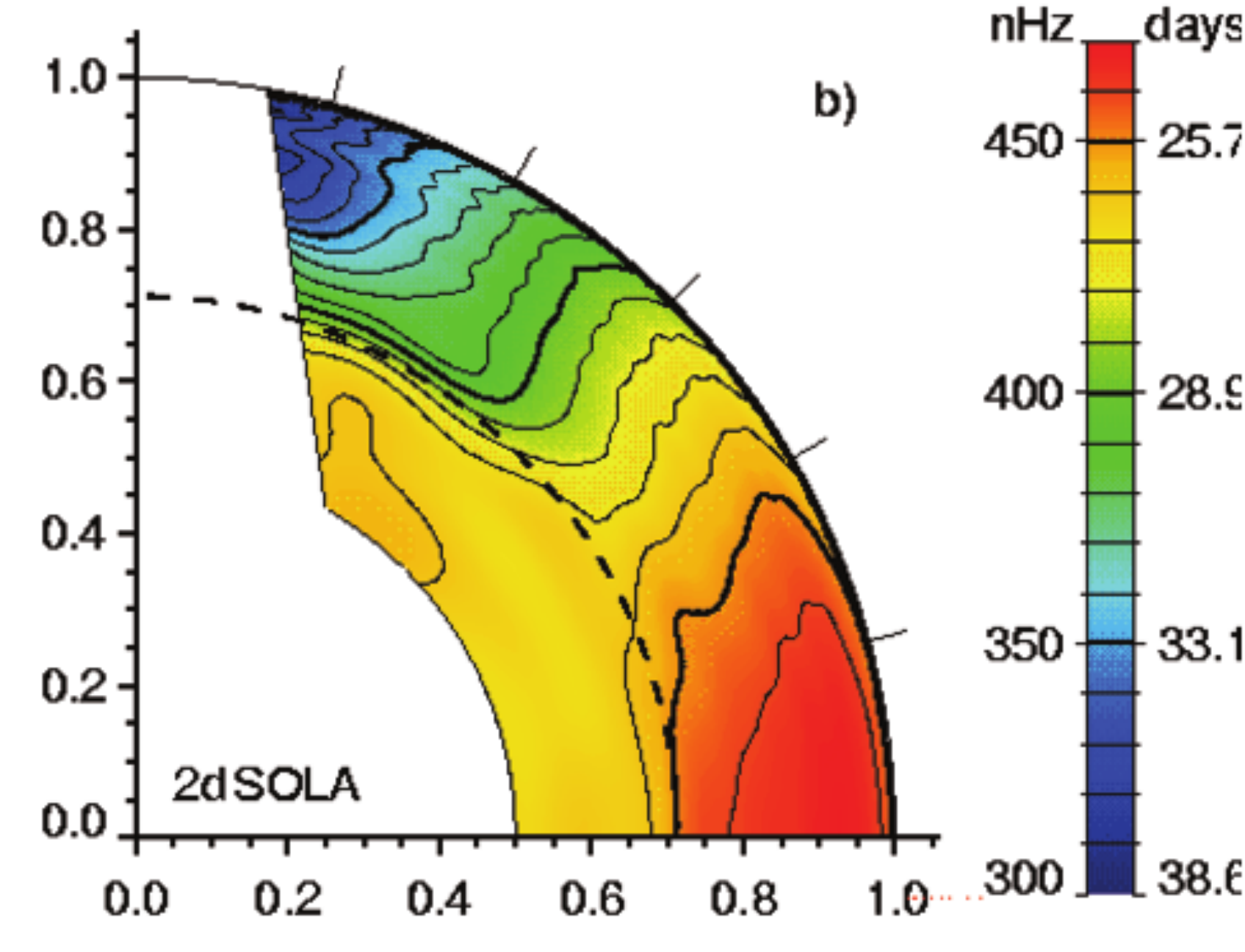}
\caption{Solar internal rotation as inferred from MDI observations. Contours of isorotation are shown. \citep[Adapted from][]{Schou1998} \label{MDIrot}}
\end{figure}

A typical rotation profile determined from helioseismology is shown in Fig.~\ref{MDIrot}. The latitudinal variation of rotation observed at the surface largely persists through the convection zone, while in the radiative interior the results are consistent with solid-body rotation. But there are two rotational shear layers, a near-surface shear layer and another shear layer termed the tachocline that is located near the base of the convection zone. 

\begin{figure}[!ht]
\centering
\includegraphics*[width=\linewidth]{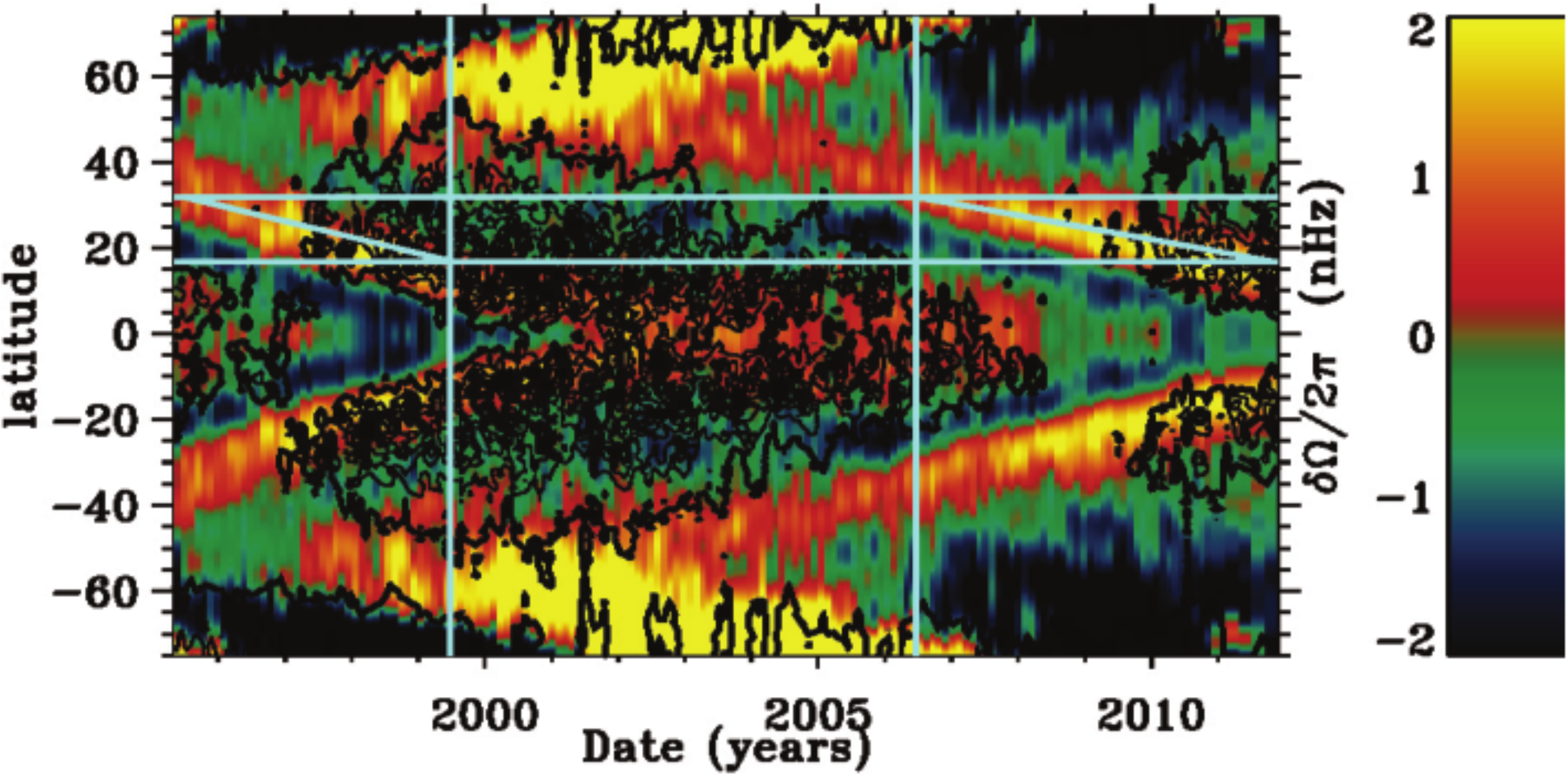}
\caption{Torsional oscillations in the form of faster (red and yellow colors) 
and slower (green and blue/black colors) zonal flows relative to the 
long-term average rotation, as inferred from helioseismology in the 
near subsurface layers of the Sun using MDI and GONG observations.
Contours of photospheric magnetic field strength are superimposed to indicate the location of contemporaneous
surface activity. 
The left vertical blue line marks the date 1997.3, when low-latitude flows most closely matched the
inference from the 2009.2 data. The vertical blue line on the right signifies 2006.4, most closely resembling the analysis from the earliest data (1996.5). The horizontal lines mark
locations of the flow bands, while the slanted lines indicate equatorward migration.
\citep[After][ courtesy of Rachel Howe.]{Howe_etal2009} \label{torsional}}
\end{figure}

Superimposed on the mean rotation profile at the surface and in the convection zone are weak but apparently coherent bands of faster and slower zonal flow 
that migrate in latitude over the course of the solar cycle (Fig.\ \ref{torsional}). These have been dubbed ``torsional oscillations'' 
\citep{Howard1980, Schou1999, Howe2000, Vorontsov2002}. The causal connection
between the flows and the magnetic manifestations of the solar cycle are 
unclear.

\subsection{Meridional circulation }\label{sec:mc-obs}

Unlike the rotation, the meridional flow does not perturb the mode
frequencies to first order.
As a consequence most measurements of the meridional flow have been
made using local helioseismology techniques, such as time-distance and
ring-diagram analysis.

\begin{figure}
\includegraphics[width=1\textwidth]{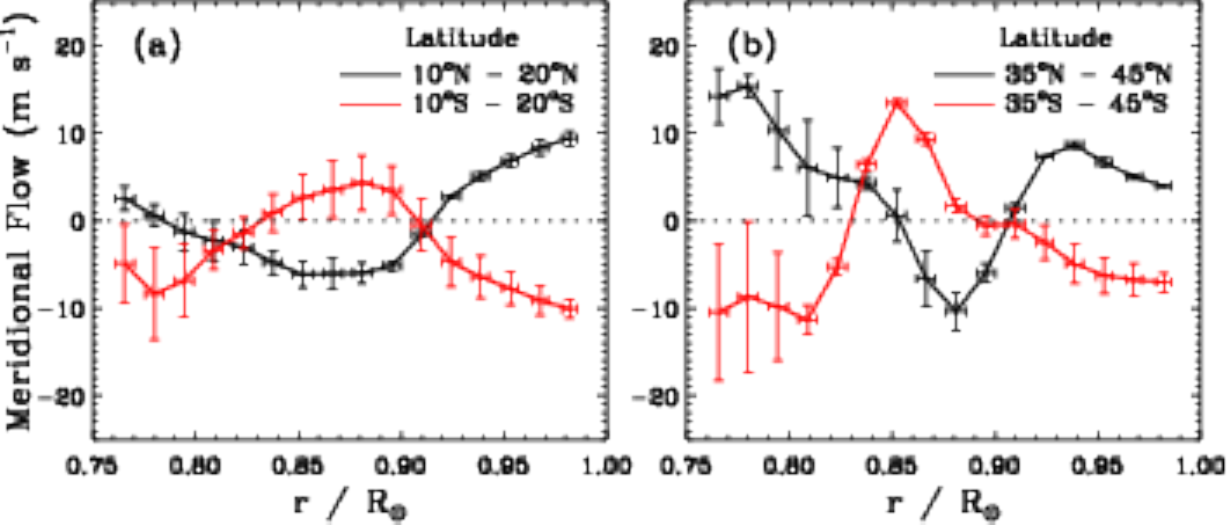}
\caption{Meridional flow profile as a function of radius at selected latitudes in
the northern (black) and southern (red) hemispheres,
obtained from time-distance helioseismology using HMI observations \citep{Zhao2013}. In the near-surface 
region the flows are poleward in both hemispheres, 
but the results indicate that 
there is a counter-cell beneath radius $0.90R_\odot$. There may be yet another cell 
beneath about $0.82R_\odot$, but the errors on the inferred flow are larger at such 
depths. (Figure courtesy of Junwei Zhao)}
\label{fig:zhao}
\end{figure}

In principle, the time-distance measurements are quite straightforward.
The N-S and S-N travel times are measured for a selection of latitude
pairs and the results inverted to obtain the meridional flow as a
function of latitude, depth and time.
In reality, and as illustrated in \cite{Zhao2012},
measurements of wave travel times (and seismic measurements in general) are prone to systematic errors,
especially when they are taken at spatial locations away from the disk center.
%\bfnote[Soften/generalize language?]
One manifestation of these errors is that the derived
flows in the interior depend on the type of observations used,
an unsatisfactory state of affairs.
Making the assumption that the error is due to a center-to-limb time
shift, Zhao et al. were able to diminish the
discrepancy between observables and obtain an estimate of the meridional
circulation over much of the solar convection zone. %\citet{Zhao2012} found
Remarkably, \citet{Zhao2013} found that the meridional circulation pattern
consists of two circulation cells in the depth direction (Figure~\ref{fig:zhao}),
 which if it were to hold up, will challenge some solar dynamo models \citep[e.g.][]{jouve07}.
While suggestions have been made regarding the origin of the
systematic errors \citep[e.g.,][]{Baldner2012},
no definite
conclusion has been reached so far.

Ring-diagram analysis has also produced a number of interesting
results on the meridional circulation 
\citep[e.g.,][]{GonzalezHernandez1999, GonzalezHernandez2006, GonzalezHernandez2008, GonzalezHernandez2010, Haber2006}.
The inferred meridional circulation exhibits temporal variations, 
including 
the occasional appearance of an equatorward cell at high
latitudes. Part of those variations
is thought to be the result of a systematic error caused by uncertainty in the Carrington elements: the inferred equatorward meridional flow at high latitutdes
shows an annual variation wiht the $B_0$ angle 
\citep{Zaatri2006}. 
Nevertheless, a multi-cell structure in latitude is also seen in the
Mount Wilson surface Doppler measurements \citep{Ulrich2010}.

\cite{lavely92} described a perturbation theoretical approach that included the study of general bulk flows. Following this theoretical framework,
\cite{rothstix99, rothstix08} studied 
the effect of giant cells and the meridional flow on the global mode 
frequencies. The effects are small, however, 
making it difficult for them 
to be measured and to be evaluated for helioseismic mappings of the 
solar interior.
Recently, new approaches concentrate on investigating the 
perturbation of the eigenfunctions rather than the eigenfrequencies when 
studying the effect of the meridional flow on solar oscillations
\citep{Woodard2000, schad11, schad12, Schad2013, woodard13}.
Since the meridional flow perturbs the solar model, 
the new oscillatory eigenstates expressed in the basis of the unperturbed 
eigenstates are mixtures of modes, which is often called mode coupling. 
Given an azimuthally symmetric meridional flow, such couplings only occur 
between modes of identical azimuthal orders $m$. Depending on the complexity of the flow when expanded in terms of Legendre polynomials as a 
function of latitude, mode coupling may occur between modes that differ in $l$
by the harmonic degree of the flow.
As a consequence, power of one mode is expected to leak into another mode and vice versa. 
The cross-spectrum allows measuring this coupling of the modes. As systematic leakage is a dominant error source in the cross-spectra due to observational and instrumental constraints, detailed knowledge of the instrument and observation conditions must be taken into account. 

\begin{figure}
\includegraphics[width=0.8\textwidth]{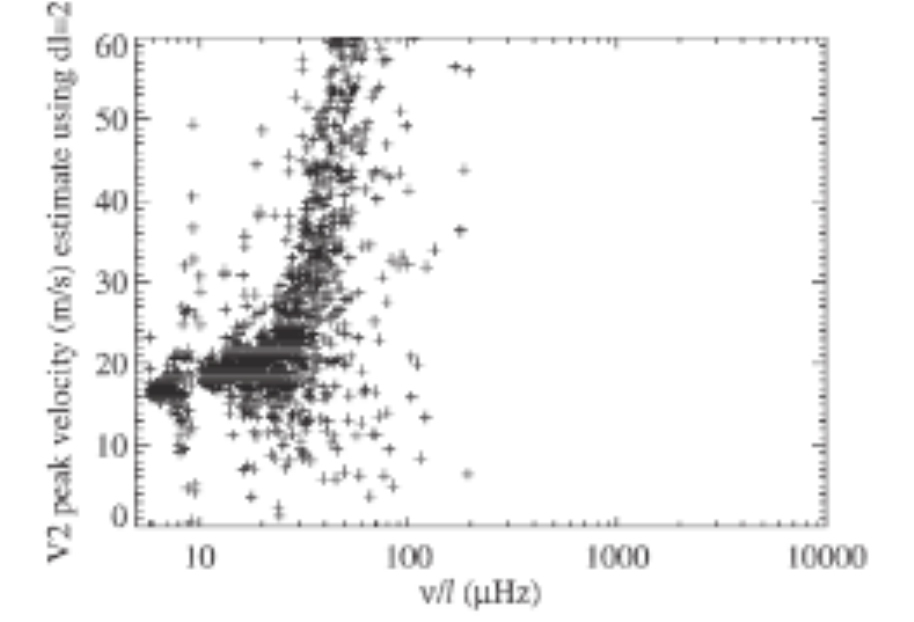}
\caption{Surface peak velocities of the horizontal meridional flow component 
of degree 2 as a function of $\nu/L$~\citep{woodard13}. 
($\nu/L$ relates directly to the radial location $r_t$ of the lower turning
point of the mode: 
e.g., for $\nu/L = 10\mu Hz$, $r_t \simeq 0.98 R_\odot$, while for 
$\nu/L = 100\mu Hz$, $r_t\simeq 0.6 R_\odot$.) 
The analysis is based on fitting a model function to the cross-spectra between p-modes and is based on 500 days of HMI data.}
\label{fig:1}
\end{figure}
  
\begin{figure}
  \includegraphics[width=0.5\textwidth]{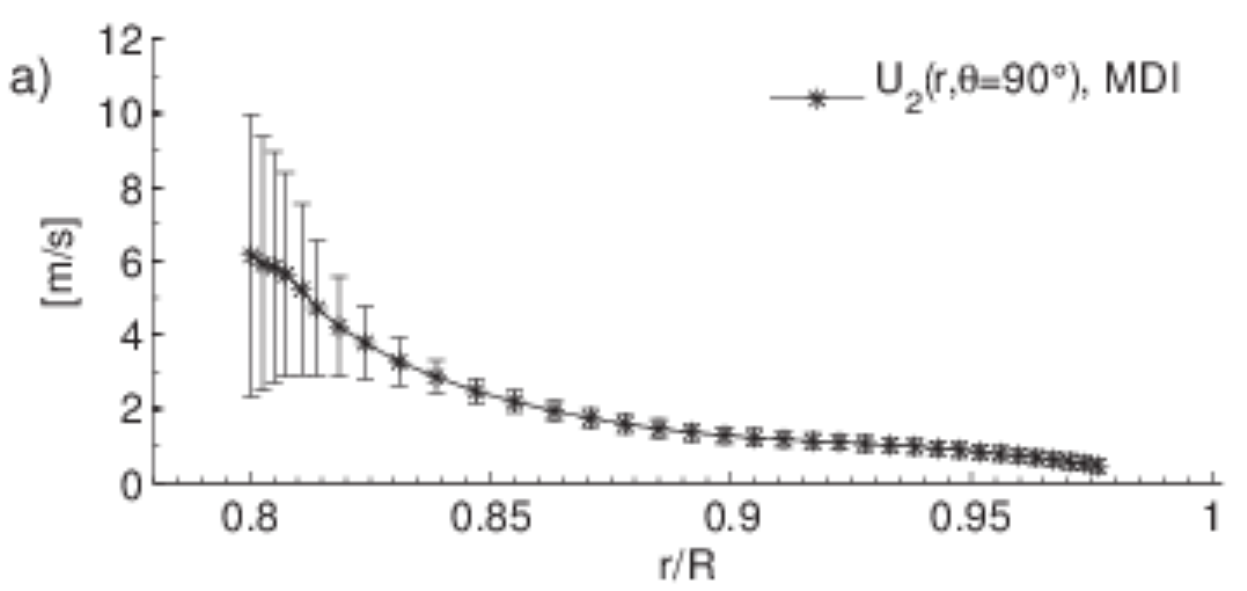}
  \includegraphics[width=0.5\textwidth]{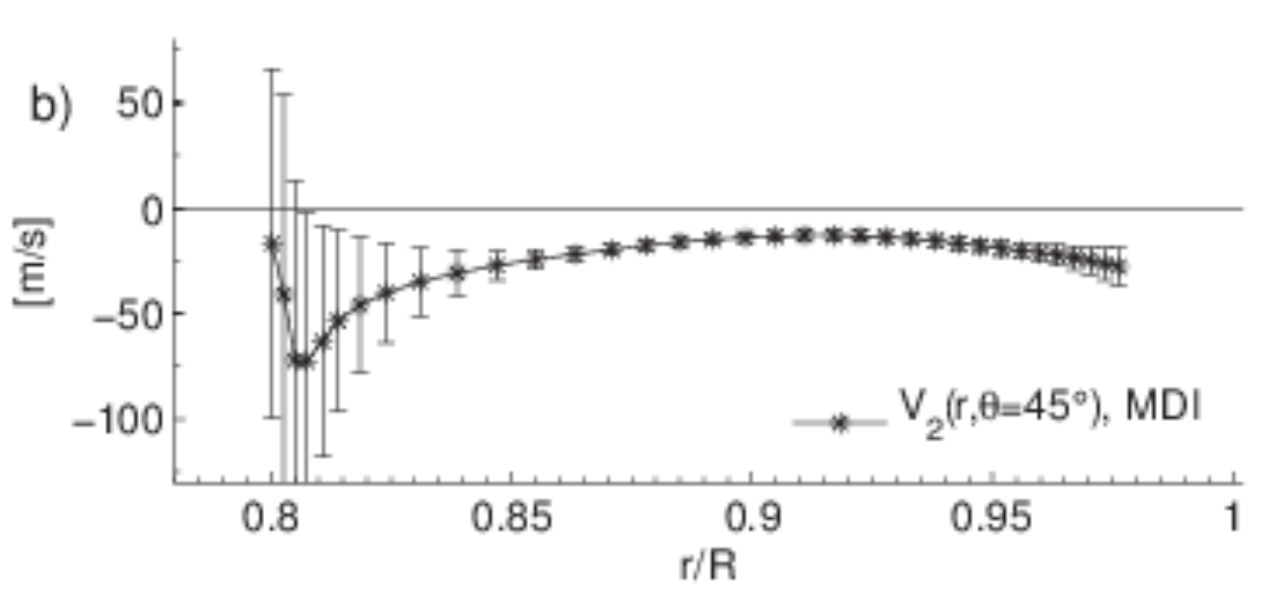}
\caption{The radial and horizontal amplitudes of the meridional flow component with harmonic degree 2 as a function of radius $r/R$ at colatitude $\theta=90^{\circ}$ (equator; left) and $\theta=45^{\circ}$ (mid-latitude; right), respectively. The flow was measured by evaluating cross-spectral amplitude ratios obtained from MDI data covering the period 2004--2010.  Positive values of $U_{2}$ indicate an upward radial flow, negative values of $V_{2}$ refer to a poleward flow~\citep{schad12}.}
\label{fig:2}       % Give a unique label
\end{figure}

Currently two approaches exist that estimate the meridional flow via 
cross-spectral analysis. 
The first one employs a fit of a model function to the observed cross-spectrum of modes that couple along the same ridge \citep{woodard13}. Figure~\ref{fig:1} shows the resulting peak velocity of the meridional flow component with a harmonic degree of 2 as a function of the ratio $\nu/(l+1/2)$, 
which is related to the inner turning point of the modes. 
Figure~\ref{fig:1} would appear to indicate a rapid increase in the meridional
flow below the surface. However, this may be an artifact related to
the center-to-limb effect suggested by \cite{Zhao2012}.
The other approach evaluates ratios of the cross-spectral amplitudes \citep{schad11, schad12} rather than a fit to the cross-spectra and includes inter-ridge couplings, too. Here the radial component of the meridional flow is measured and the horizontal meridional flow is obtained via mass conservation. Through the use of ratios, 
this method appears robust against systematic errors present in the cross-spectra. Figure~\ref{fig:2} displays the result for the radial meridional flow measured at the equator and the horizontal meridional flow measured at $45^\circ$ latitude as a function of fractional solar radius. Again, the harmonic degree of the flow component is 2.

\begin{figure}
  \includegraphics[width=0.93\textwidth]{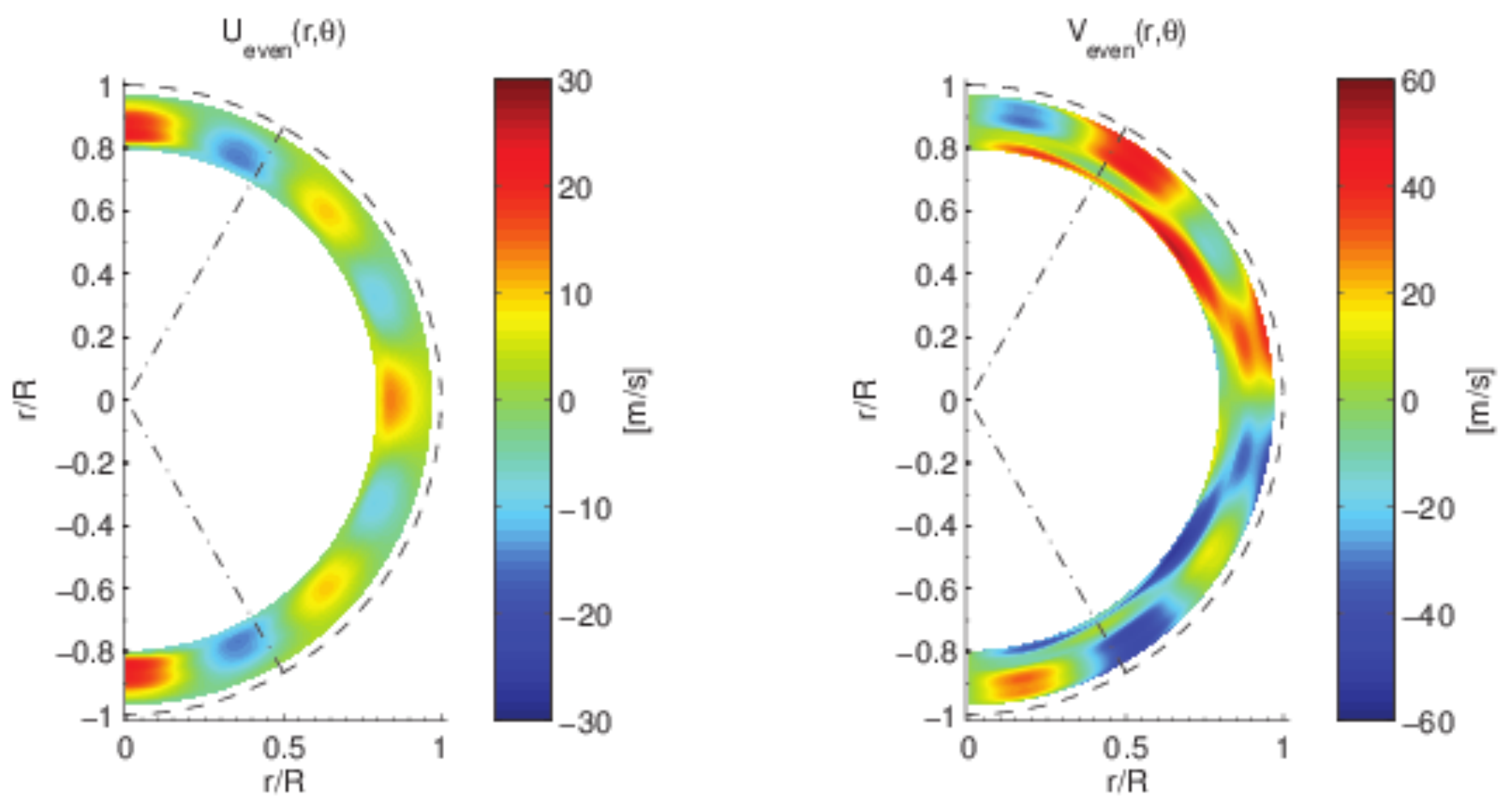}\\
  \includegraphics[width=0.93\textwidth]{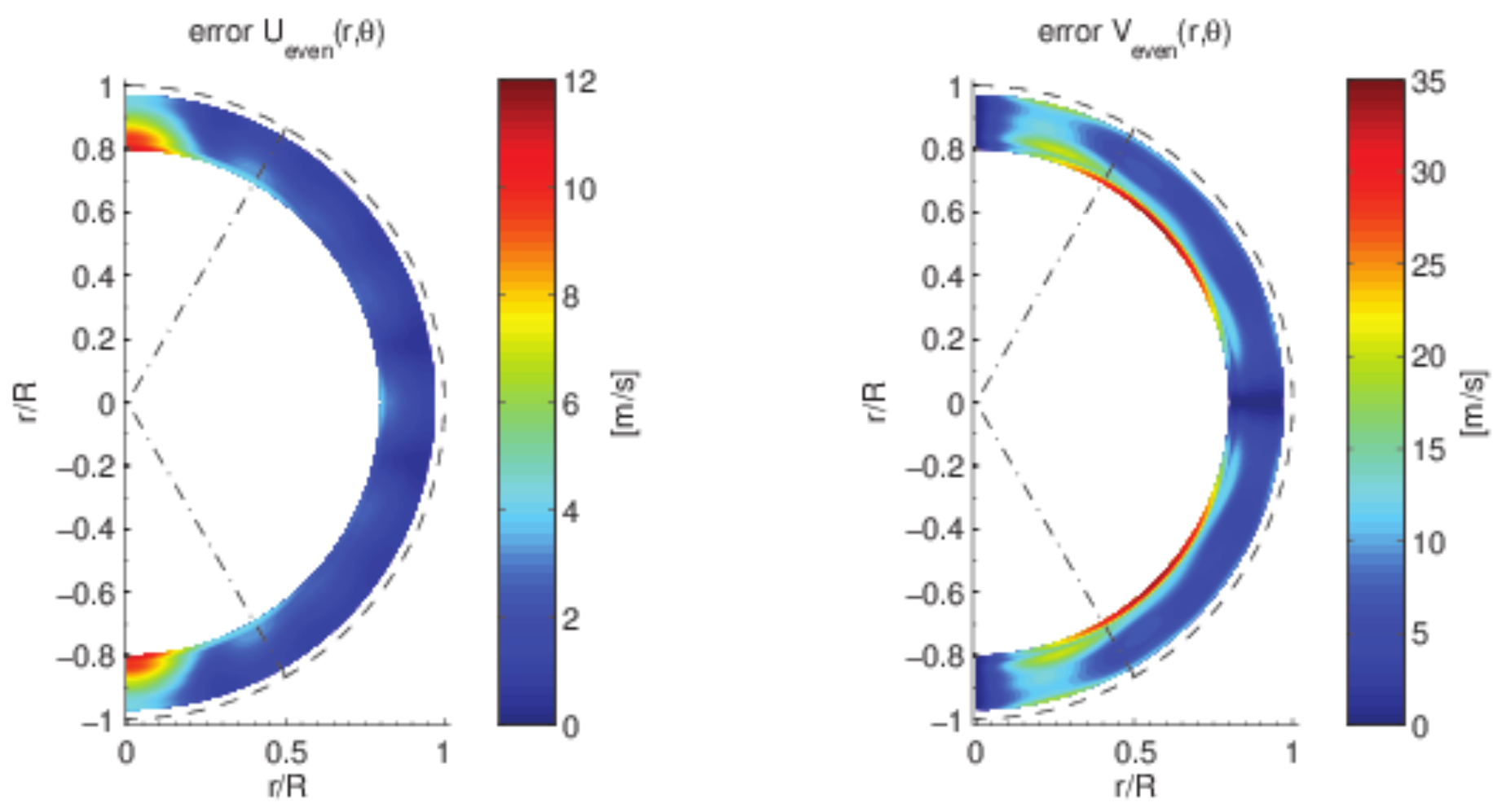}
\caption{
Cross-sections through the meridional flow composed of the harmonic degrees 2, 4, 6, 8 (top) as a function of fractional radius and latitude based on evaluating cross-spectral amplitude ratios obtained from MDI data covering the period 2004--2010~\citep{Schad2013}. The dashed-dotted lines mark the latitudes of $60^\circ$. The radial flow is displayed on the left where positive (negative) values correspond
to outward (inward) directed flows. The horizontal flow is displayed on the right; positive (negative) values correspond to northward (southward) directed flows. The 1$\sigma$ standard error of
the composite flow is given in the lower panels.}
\label{fig:2b}       % Give a unique label
\end{figure}

Evaluating mode couplings caused from flow components with a harmonic degree up to 10, \citet{Schad2013} were able to measure some of these components down to a fractional solar radius of approximately 0.5. Only the flow components with harmonic degrees 2 and 8 deviate significantly from zero, and give a first hint on a meridional flow that is confined between the tachocline region and the solar surface and that exhibits a multi-cellular structure as a function of radius and latitude. Figure~\ref{fig:2b} shows a composite of the even meridional flow components.

With both techniques the inferred solar near-surface horizontal flow 
is consistent with subsurface flow measurements from local helioseismology, 
with a poleward directed flow with a peak amplitude of order 20 m/s. 
However, we note that at greater depths the results shown in Figures~\ref{fig:zhao}
and those derived by \cite{Hathaway2012} are inconsistent with Figure~\ref{fig:2b}. Both \citet{Zhao2013}
and \citet{Hathaway2012} infer a return (equatorward) flow at $r \approx 0.9R_\odot$ whereas
Figure~\ref{fig:2b} suggests a flow that remains poleward up till a depth of $r=0.8R_\odot$.  
Further, although qualitatively similar up to a depth of $r = 0.9R_\odot$ or so, the amplitudes of the return flow 
inferred by \citet{Hathaway2012} and \citet{Zhao2013} are quantitatively different.
Some of these discrepancies could be attributed to the fact that the analyzed data sets cover different time frames. Nevertheless, this is indicating that the results should be
treated with some caution and that further work is needed to determine the 
source of the discrepancies. 
%\bfnote[What are the inconsistencies? Are Zhao et al. and Fig. 6 consistent?]

\subsection{Convection and overshoot }\label{sec:con-ov}

\begin{figure}[!ht]
\centering
\includegraphics*[width=\linewidth]{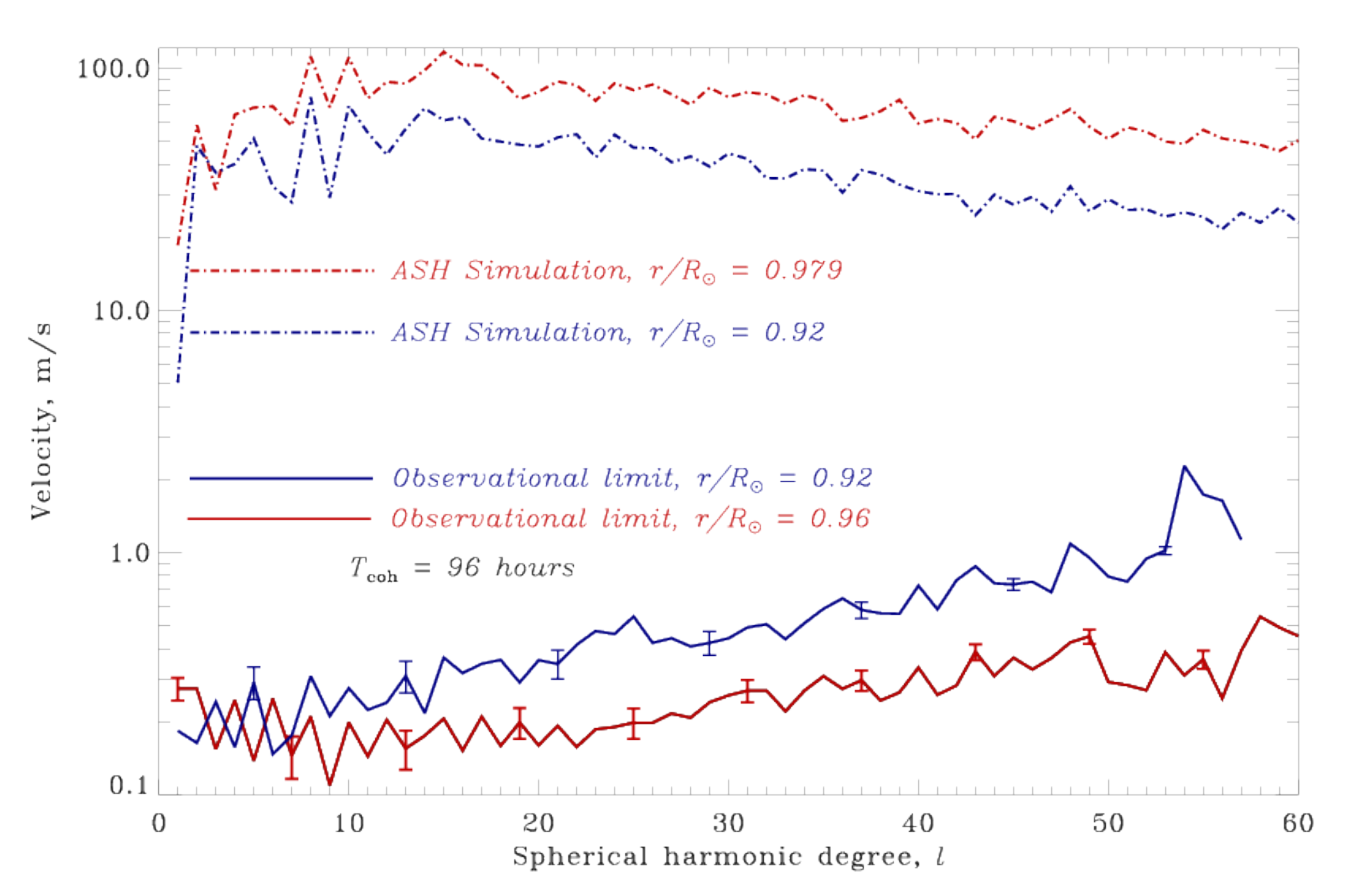}
\caption{Seismic constraints obtained by \citet{Hanasoge12} using data from HMI \citep{hmi}. Assuming a convective coherence time of 96 hours,
\citet{Hanasoge12} obtained upper bounds on the observed convective spectrum. Shown in comparison is the Anelastic
Spherical Harmonic (ASH) convective spectrum. These differences suggest that convection in the Sun may be operating in a strongly non-MLT regime. \label{constraints}}
\end{figure}

Using techniques of time-distance helioseismology \citep{duvall, duvall03, gizon2010}, \citet{Hanasoge12} placed stringent bounds on the interior
convective velocity spectrum. Two-point correlations measured 
from finite temporal segments of length of the observed line-of-sight photospheric Doppler velocities, taken by HMI, 
were used in the analysis. 
These correlations were spatially averaged according to a deep-focusing geometry \citep{Hanasoge10} in order to image the 
interior. Convective coherence timescales \citep{spruit74, gough77, miesch_etal_08} were taken into account in choosing the temporal length of the data. 
By construction, these measurements are sensitive to the 3 components of the underlying flowfield, i.e., longitudinal, latitudinal or radial, 
at specific depths of the solar interior ($r/R_\odot = 0.92, 0.96$). The constraints are shown in Figure~\ref{constraints}.
The stark difference between observations and simulations suggests that the convection in the Sun may be operating in a strongly
non-MLT regime. A plume-based thermal transport mechanism in this alternative regime has been the subject of speculation by, e.g., \citet{spruit97}
and was explored further by \citet{rempel2005}. The alternate mechanism put forward by \citet{spruit97} would be able to account for the
outward thermal transport of a solar luminosity's worth of heat flux at extremely low flow speed.
However, the transport of angular momentum, i.e., the maintenance of differential rotation and meridional
circulation, is not as easily explained. Based on scaling arguments, \citet{Miesch12} estimate the minimum convective kinetic
energy (and associated Reynolds' stresses) required to sustain these large-scale flow circulations. 
\citet{gizon12} provide an interesting comparison between seismology, simulation, and the phenomenology of \citet{Miesch12}.

Just as convection leads to an essentially adiabatically stratified envelope in
the outer thirty per cent of the Sun, convective overshooting is expected to 
modify the stratification of the region in which it takes place 
\citep{Skaley1991, Deng2008}.
In the simplest picture, overshoot at the base of the convection zone leads
to an extension of the adiabatically stratified region, with a more-or-less
abrupt transition beneath that to the subadiabatic stratification of the 
radiative interior. Such a signature of a sharp transition in the sound-speed
profile of the Sun has been sought with helioseismology
\citep{Basu1994, Monteiro1994, Roxburgh1994, jcd1995} 
In principle, the amplitude of the abrupt transition can then be used to infer the extent of 
the overshoot region, and upper bounds have been quoted on the extent of the 
overshooting of about $0.005R_\odot$ \citep{Monteiro1994}.

In fact, it appears that the transition between the convection zone and the 
radiative interior in the Sun is actually smoother even than in solar models
that have no convective overshooting, which is a challenge to the simple 
picture above. A different model of convective overshoot has been 
proposed by \cite{Rempel2004} wherein the overshoot is modeled with 
discrete plumes with a spectrum of strengths and hence depths of penetration
into the radiative interior \citep[also see][for models of overshoot]{zahn91}. Depending on the spectrum adopted, this can 
give a smoother transition, indeed the subadiabatic stratification can 
occur even towards the bottom of the convective region. 
\cite{jcd2011} investigated the seismic signature of such models,
compared with helioseismic observations, and found that some models of this 
class fitted the observations better 
than the simpler models without (the ``Standard solar model'') or 
with overshoot. They concluded that overshoot is necessary to improve the
agreement between models and helioseismic constraints, that the required
stratification profiles are outside the realm of classic ``ballistic'' 
overshoot models, and that the lower part of the convection zone is 
likely substantially subadiabatic. Overshoot has also been studied in detail
using numerical simulations, with local regions \citep{toomre02,rogers05} and
global 3-D domains \citep{brun11}.

\section{Advances in modeling the Sun's internal dynamics}
\label{sec:4}
Having discussed the observational findings from helioseismology about the
Sun's internal dynamics, we consider now what is understood from theory and 
numerical simulations.
\subsection{Solar convection and mean flows}\label{sec:mean-flows}
Recent insights into the nature of global-scale solar convection and
mean flows have centered around two simple but powerful concepts.
The first is thermal wind balance, which expresses the force balance 
in the meridional plane between the inertia of the differential
rotation (Coriolis/centrifugal terms) and thermal gradients in
latitude (baroclinic term):
\begin{equation}\label{eq:twb}
\frac{\partial \Omega^2}{\partial z} = \frac{g}{r \lambda C_P} ~ 
\frac{\partial \left<S\right>}{\partial \theta} ~~~.
\end{equation}
Here we use a mixture of spherical polar coordinates $(r, \theta, \phi)$
and cylindrical coordinates $(\lambda, \phi, z)$, with $S$
denoting the specific entropy.
Angular brackets denote averages over longitude and time while $g$ and $C_P$ are
the gravitational acceleration and the specific heat at constant pressure.
Equation (\ref{eq:twb}) holds if the meridional components of the convective
Reynolds stress, the Lorentz force, and the viscous diffusion are small 
relative to the Coriolis and baroclinic terms.   This is supported by
a diverse range of theoretical and numerical modeling efforts and
points to the central role of baroclinicity in accounting for the
conical nature of the $\Omega$ isosurfaces inferred from helioseismic
rotational inversions, as seen in Fig.\ \ref{MDIrot}
 \citep{Kitchatinov95,Elliott00,Robinson01,Rempel05,Miesch06,Balbus09}.

The second concept is that of gyroscopic pumping, which can be expressed
by the following balance in the zonal component of the momentum equation:
\begin{equation}\label{eq:gp}
\left<\rho {\bf v}_m\right> \mbox{\boldmath $\cdot \nabla$} {\cal L} = {\cal F}
\end{equation}
where ${\bf v}_m$ denotes the meridional components of the
velocity, and ${\cal L} = \lambda^2 \Omega$ is the specific angular momentum.
The right-hand side of eq.\ (\ref{eq:gp}) is expressed as a generalized torque
${\cal F}$ but it can be loosely regarded as the negative divergence of the 
convective Reynolds stress.  Lorentz forces and viscous torques may also 
contribute to ${\cal F}$ but they are unlikely to be the central mechanism
for establishing the solar differential rotation.  For a derivation of
eq.\ (\ref{eq:gp}) and a thorough discussion of its implications, see
\cite{Miesch11}.

Together these two concepts, reflected by the dynamical balances in
equations (\ref{eq:twb}) and (\ref{eq:gp}), provide a theoretical
foundation for interpreting helioseismic inversions and numerical
models.  For example, by making the additional ansatz that $\Omega$
and $\left<S\right>$ isosurfaces coincide, Balbus and colleagues have
shown that solutions to equation (\ref{eq:twb}) coincide remarkably
well with helioseismic rotational inversions.  However, there are two
caveats to this result.  First, it does not explain why the solar
equator rotates faster than the poles; adding an arbitrary cylindrical
angular velocity component $\Omega^\prime(\lambda)$ to $\Omega$
leaves equation (\ref{eq:twb}) unchanged (geostrophic degeneracy) 
so a given $\left<S\right>$ profile is consistent with an infinite 
number of $\Omega$ profiles, some solar-like (fast equator, slow poles), 
some anti-solar (slow equator, fast poles).  Second, a compelling 
theoretical justification of why $\Omega$ and $\left<S\right>$ surfaces 
should coincide remains an outstanding challenge 
\citep[though see][for one perspective]{Balbus12}.

%By considering the time-dependent analogues of equations (\ref{eq:twb})
%and (\ref{eq:gp}), \cite{Miesch12} have argued that thermal gradients
%(baroclicity) alone cannot account for either the meridional circulation 
%or the differential rotation inferred from helioseismic inversions.
%Rather, the main driver for both must be the convective angular 
%momentum transport as embodied in the torque term ${\cal F}$.
%\cite{Miesch12} then exploit this result to set bounds on the 
%amplitude of convection $V_c$ needed to sustain the mean flows 
%inferred from helioseismology.  Their result, $V_c \gtrsim$ 30 m s$^{-1}$ 
%in the upper convection zone, is comparable to estimates from 
%convection simulations and mean-field models but is somewhat
%larger than the lower limits obtained by \cite{Hanasoge10,Hanasoge12} 
%based on time-distance inversions for spherical harmonic modes of degree
%$\ell \lesssim 60$.  This suggests that the structure, evolution, and
%detection of giant cells may be more subtle than expected.  Resolving 
%this puzzle is a major challenge for future convection models 
%and helioseismic investigations.

\begin{figure*}
  \includegraphics[width=\textwidth]{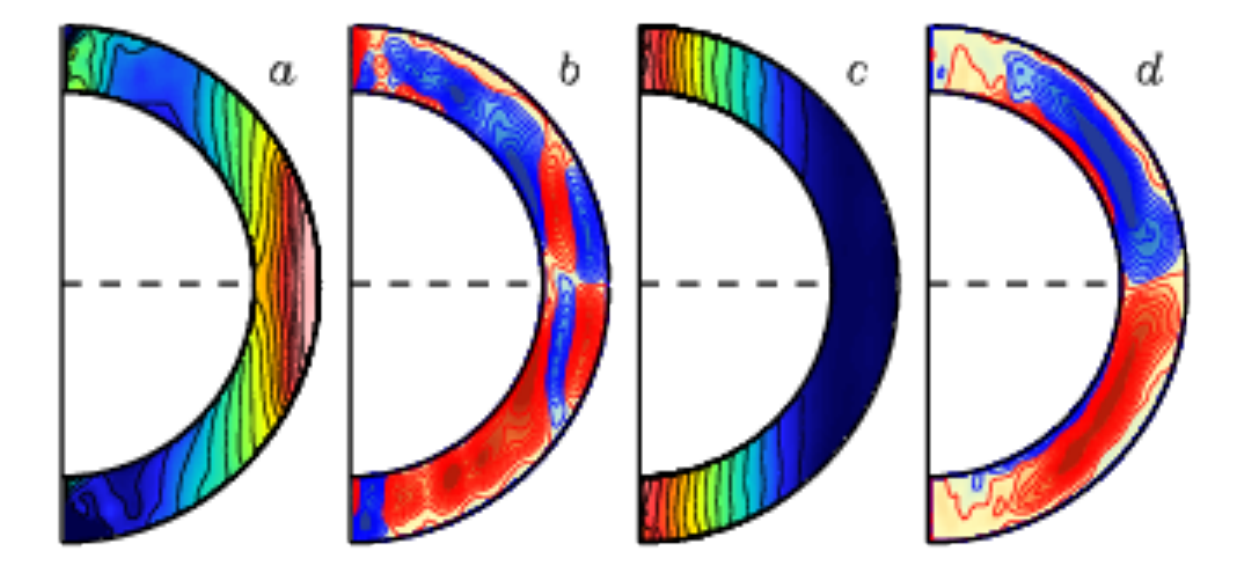}
\caption{Mean flow regimes in spherical convection, from
\cite{Featherstone14}.  Shown are the differential rotation
($a$, $c$) and meridional circulation profiles ($b$,$d$)}
in two simulations of global solar convection.  Pink/yellow
and blue/black tones denote faster and slower rotation in 
frames $a$,$c$ while red and blue denote clockwise and 
counter-clockwise circulation in $b$, $d$.  On the left
($a$,$b$) is the rapidly-rotating regime characterized by
a solar-like $\Omega$ profile and multi-celled meridional
circulation.  On the right ($c$,$d$) is the slowly-rotating regime
characterized by an anti-solar differential rotation and
and a single dominant circulation cell per hemisphere.
The Sun may be near the transition.
\label{fig:cfig}
\end{figure*}

A puzzle that has received much attention
recently is that of the subsurface structure of the meridional
circulation.  It has been realized since the pioneering work of
\cite{Gilman77} that spherical convection simulations exhibit two
rotation regimes, delineated by the Rossby number $R_o = V_c /
(2\Omega L_c)$ where $V_c$ and $L_c$ are characteristic velocity and
length scales for the convection.  As $R_o$ is increased across values
of order unity, the differential rotation undergoes a transition from
being solar-like to anti-solar, as illustrated in Fig.\
\ref{fig:cfig}.  More recent modeling efforts are clarifying this
transition and assessing its implications for the meridional
circulation 
\citep{Kapyla11,Kapyla14,Gastine13,Gastine14,Guerrero13,Hotta14a,Hotta14b,Featherstone14}.  
The fast-rotating regime generally exhibits multiple-cell profiles while
the slow-rotating regime exhibits circulation profiles dominated by a
single cell per hemisphere (Fig.\ \ref{fig:cfig}).  This transition
can be understood in terms of a shift in the nature of the convective
Reynolds stress ${\cal F}$ as discussed above in connection with 
eq.\ (\ref{eq:gp}).
The Sun is likely near the transition so it is unclear which meridional
flow regime it may be in \citep{Featherstone14}.

%%%%%%%%%%%%%%%%%%%%%%%%%%%%%%%%%%%%%%%%%%%%%%%%%%%%%%%%
 
Gyroscopic pumping relies on the zonal component of the
convective Reynolds stress, which induces meridional circulation
by means of the Coriolis force.  The meridional components of the
Reynolds stress and Lorentz force can also establish meridional
circulation by breaking the thermal wind balance (TWB; eq.\ \ref{eq:twb}).
This occurs in many mean-field models that represent meridional
Reynolds stresses as a turbulent diffusion and solve for the
steady flow profiles \citep{Kitchatinov12,Dikpati14}\footnote{Though
see \cite{Rempel05} for an example of a time-dependent mean field model
in which gyroscopic pumping is the dominant meridional flow driver.}.  
Departures from TWB occur particularly in the boundary layers, which 
can exert a disproportionate influence on the global meridional
circulation profile.  In particular, the shallow equatorward return
flow inferred from recent helioseismic inversions and photospheric 
feature tracking (section \ref{sec:mc-obs}) may be a boundary layer
phenomenon, reflecting the penetration depth of surface-driven
convective plumes \citep{Featherstone14}.

%%%%%%%%%%%%%%%%%%%%%%%%%%%%%%%%%%%%%%%%%%%%%%%%%%%%%%%%

Other current puzzles that are actively being investigated include 
the origin of the thermal gradients necessary for TWB
(eq.\ \ref{eq:twb}) and the nature of the near-surface shear layer (NSSL).  
Proposed mechanisms for the former include the influence of rotation 
on convective heat transport \citep{Kitchatinov95,Brun02,Kapyla11}, 
the influence of rotational shear on convective heat transport 
\citep{Balbus09,Balbus12}, and thermal coupling to the subadiabatic 
tachocline \citep{Rempel05,Miesch06}.  For recent perspectives
on the latter puzzle (the NSSL), see \cite{Miesch11}, \cite{Gastine13},
\cite{Guerrero13}, and \cite{Hotta14b}.

\subsection{Magnetoconvection and surface magnetism}

Magnetoconvective processes, i.e., the interaction between magnetic
field and convective flows, play a central role in the generation,
intensification, transport, and dissipation of magnetic flux in the
convection zone and in the lower atmosphere of the Sun. Significant
progress in our understanding of these processes has been brought about
during the last decade by the combination of high-resolution
observations and sophisticated numerical simulations \citep[for recent
reviews of various aspects of solar magnetoconvection,
see][]{Miesch:2005, Stein:2012, Weiss:2012,
Schuessler:2013}. Prominent examples are the formation of intergranular
magnetic flux concentrations \citep{Bercik:etal:1998, Voegler:etal:2005,
Stein:Nordlund:2006, Schaffenberger:etal:2006} and magnetized vortices
\citep{Voegler:2004a, Shelyag:etal:2011, Moll:etal:2012,
Wedemeyer:etal:2012a, Kitiashvili:etal:2012, Shelyag2013}, 
the near-surface structure
and dynamics of sunspot umbrae and penumbrae
\citep{Schuessler:Voegler:2006, Heinemann:etal:2007, Rempel:etal:2009,
 Rempel:etal:2009a, Rempel:2011a, Rempel:2011b, Rempel:2012a}, the emergence of active
regions \citep{Cheung:etal:2007, Martinez:etal:2008,
Cheung:etal:2010}, as well as small-scale dynamo action in the deep
convection zone \citep{Brun:etal:2004} and in its near-surface layers
\citep{Voegler:Schuessler:2007, Graham:etal:2010}.

Recent developments concern magnetoconvection simulations in
computational boxes of down to 50~Mm depth, so that the effect of 
flows at supergranular scales can be
studied. These flow patterns imprint their signature on the
distribution of magnetic flux at the visible surface, leading to
mesogranular and supergranular network patterns. If a sufficient
amount of vertical background flux is present, bigger flux
concentrations resembling observations of dark pores are formed by
flux expulsion and suppression of convective energy transport. An
important requirement for this kind of simulations is a sufficiently
big aspect ratio of the computational box (ratio of horizontal size to
depth): since the largest flow structures typically have a horizontal
extension of the order of the box depth, the horizontal size of the
box should be at least 3-4 times its depth, so that a sufficiently
large number of cells is present, thus minimizing the effect of the
(typically periodic) side boundaries.  Results based on simulations
with aspect ratios of unity \citep[e.g.,][]{Kitiashvili:etal:2010}
should therefore be considered with some
caution. 

\begin{figure}[ht!]
% \vspace*{0 cm}
\begin{center}
 \includegraphics[width=0.45\linewidth]{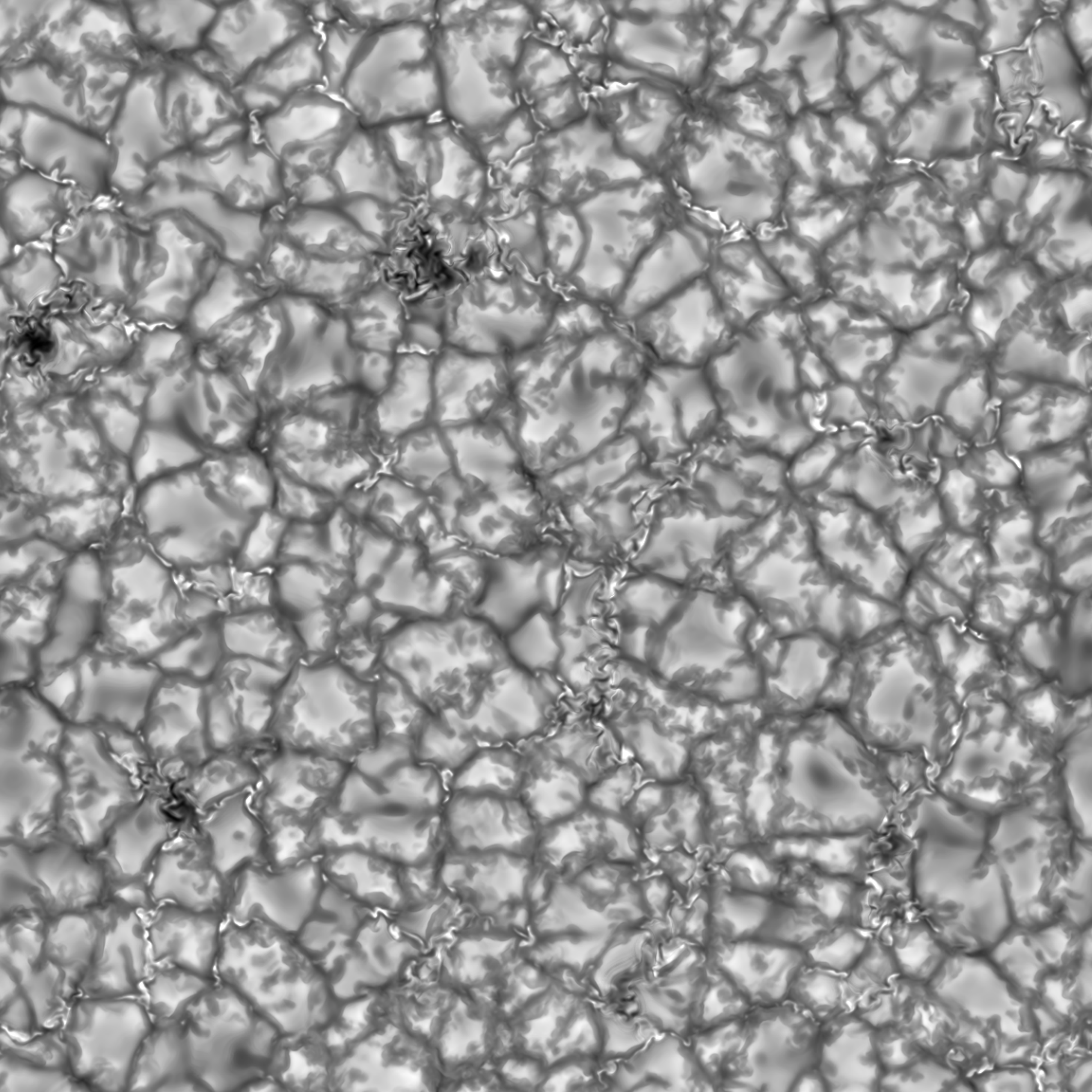}\hspace{2mm}
 \includegraphics[width=0.45\linewidth]{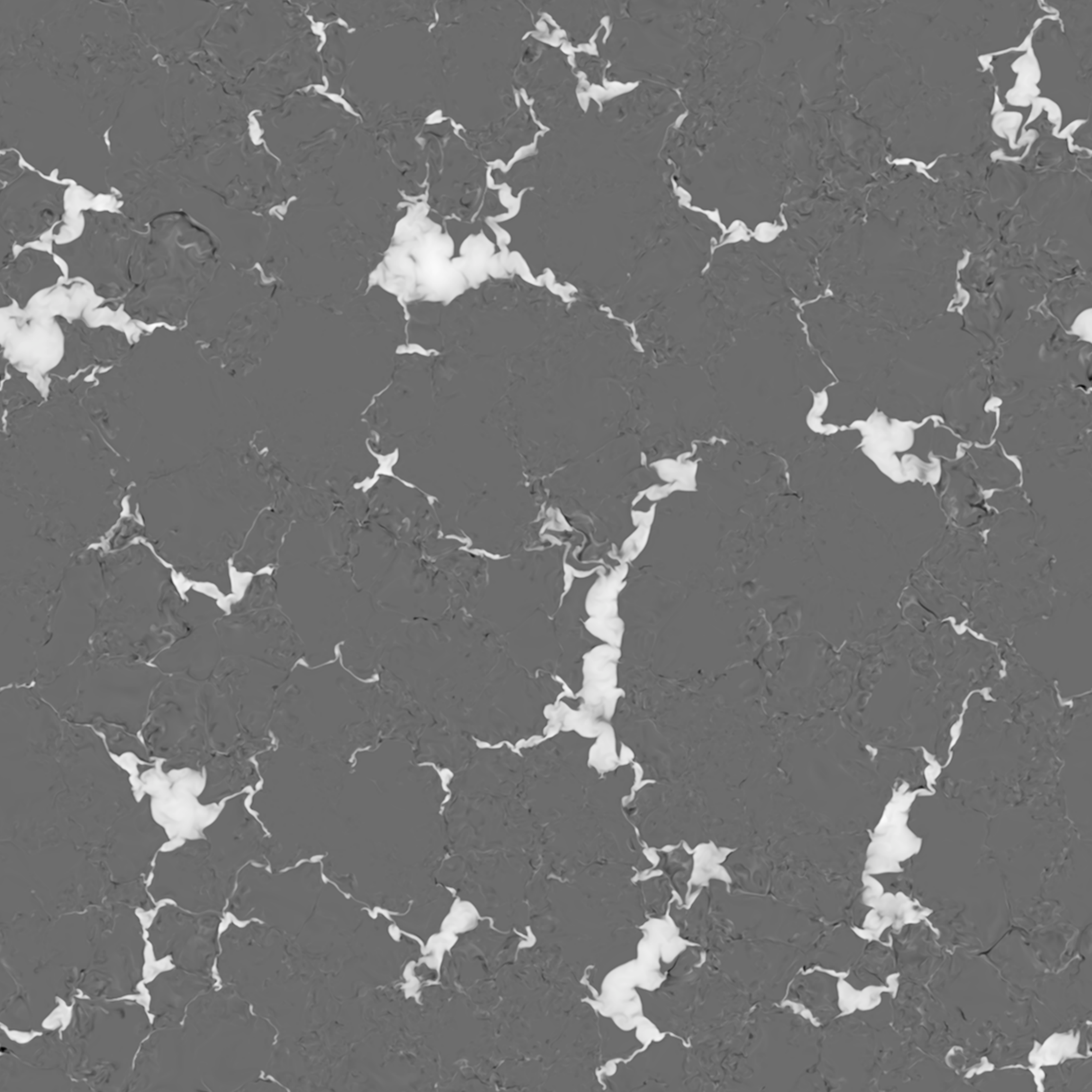}\vspace{3mm}\\
 \includegraphics[width=0.45\linewidth]{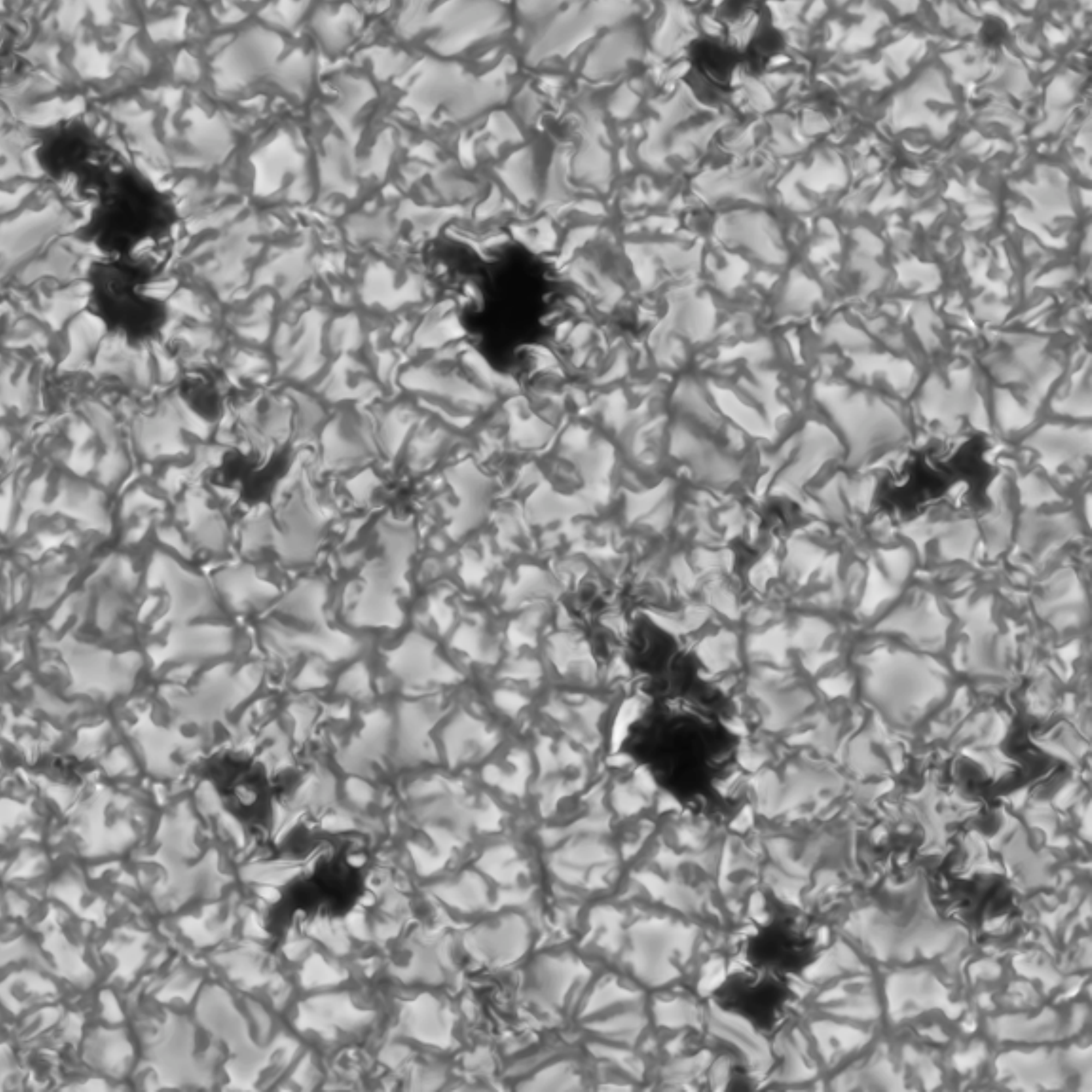}\hspace{2mm}
 \includegraphics[width=0.45\linewidth]{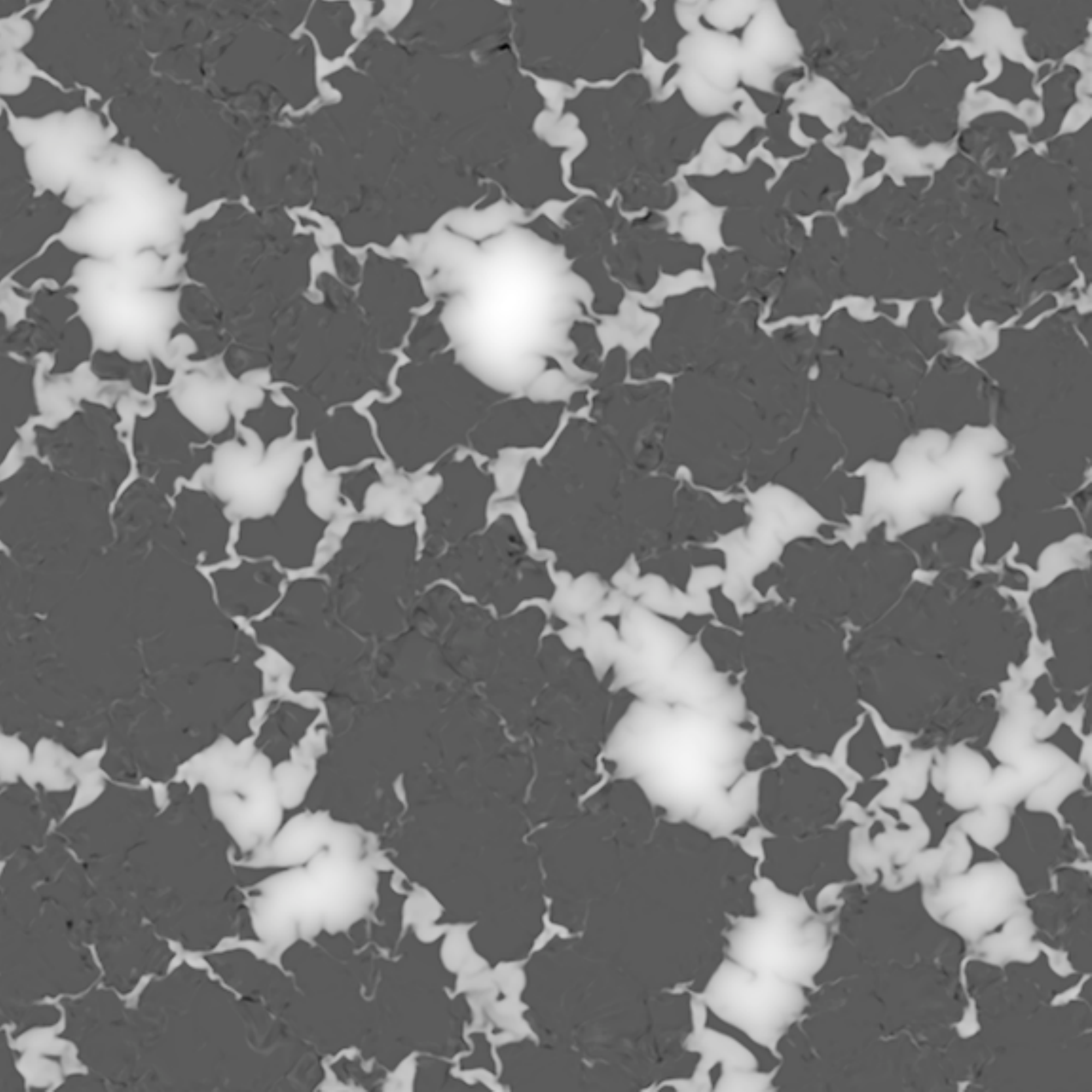}
%\vspace*{0 cm}
 \caption{Maps of bolometric brightness (left panels) and vertical
   magnetic field at the optical surface (right panels; dark grey
   represents very weak field) from magnetoconvection simulations with
   average vertical background fields of $100\,$G (upper row) and
   $400\,$G (lower row). The horizontal extension of the computational
   box was $24\,{\rm Mm}\times24\,{\rm Mm}$ and the depth $6\,{\rm
     Mm}$. The bigger amount of magnetic flux in the case with
   $400\,$G background field leads to the formation of proper pores
   while only micropores are present in the case with $100\,$G. The
   latter case has a twice higher horizontal grid resolution
   ($20.8\,$km), which leads to a better representation of the
   small-scale bright flux concentrations in the intergranular lanes.
 }
\label{fig_msch_1}
\end{center}
\end{figure}

Fig.~\ref{fig_msch_1} shows magnetic flux concentrations and pore-like
structures forming in simulations in a box of $24\,{\rm
  Mm}\times24\,{\rm Mm}$ horizontal size and $6\,{\rm Mm}$ depth. The
simulations differ by the amount of imposed vertical background flux:
in the case with a horizontally averaged vertical field of $\langle
B_z\rangle=100\,$G (upper panels), a patchy network on mesogranular
scales and a few micropores formed; for $\langle B_z\rangle=400\,$G
(lower panels), the network is more pronounced and a number of dark
pores of the size of a few granules are present. Note that $\langle
B_z\rangle=400\,$G is already at the high end for solar plage areas;
consequently we do expect even bigger structures (sunspots) to form
spontaneously from existing background flux. This is consistent with
the well-known observational fact that sunspots invariably form in the
course of flux emergence and never from pre-existing flux in a mature
plage region.

\begin{figure}[ht!]
% \vspace*{0 cm}
\begin{center}
 \includegraphics[width=\linewidth]{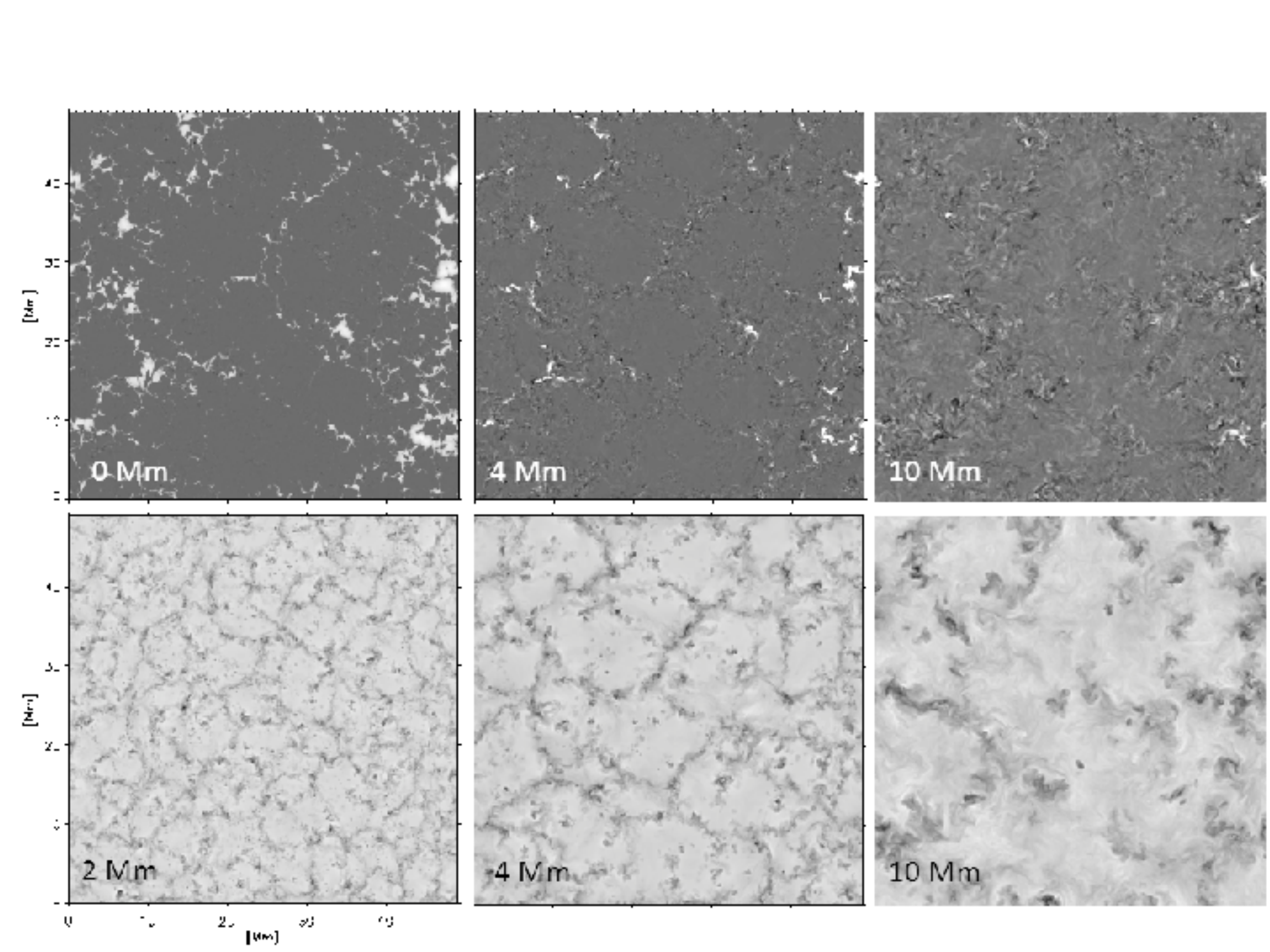}
%\vspace*{0 cm}
 \caption{Maps of the vertical magnetic field (upper panels; dark grey
   represents very weak field) and vertical flow velocity (lower
   panels; darker shades represent downflows, brighter shades upflows)
   at various depths below the optical surface from a
   magnetoconvection simulation with average vertical background
   fields of $100\,$G (courtesy of M.C.M.~Cheung). The horizontal
   extension of the computational box was $49.2\,{\rm
     Mm}\times49.2\,{\rm Mm}$ and the depth $15.4\,{\rm Mm}$. The
   spatial patterns of the observable surface field (upper left panel)
   reflect the horizontal scales covered by the flow
   patterns in the depth range of the simulation.  }
\label{fig_msch_2}
\end{center}
\end{figure}

Figure~\ref{fig_msch_2} shows a comparison of flows and magnetic field
patterns for various depths. The corresponding magnetoconvection
simulation ($\langle B_z\rangle=100\,$G) was carried out by
M.C.M. Cheung in a computational box with $49.2\,{\rm
  Mm}\times49.2\,{\rm Mm}$ horizontal size and reaching down to about
14~Mm below the optical surface. The various vertical flow patterns
present at the different depths up to about supergranular scale are
reflected in the multi-cellular distribution of vertical magnetic flux
at the surface, owing to flux expulsion by the corresponding
horizontal flows patterns. This offers the possibility to compare the
simulation results with actual observations. Since the average
properties of the simulated convection (e.g., depth profiles of
horizontally averaged thermodynamic quantities, horizontal scales and
velocities) are in good agreement with mixing-length models, this
would also provide a consistency check for such models and shed light
on the question whether solar convection might actually work in a
completely different, essentially unmixed regime with cool, narrow
downdrafts traversing the whole convection zone and driving a very
slow, broad upflow \citep{spruit97}. Such a regime appears to be
favored by recent indications from helioseismology
(Fig.~\ref{constraints}). 
Such `slow', unmixed solar convection would have severe consequences for
models of the solar dynamo and for the generation of differential
rotation and meridional flow 
\citep[][see also the discussion in Section \ref{sec:con-ov} above]{Miesch:etal:2012}. 
Existing numerical simulations invariably show well mixed convection with
velocities consistent with mixing-length models. This is the case for
anelastic simulations of convection in the deep convection zone
\citep[e.g.,][]{Miesch:2005} as well as for compressible simulations
in the upper layers of the convection zone, which include the driving
by radiative losses in the photosphere
\citep[e.g.,][]{Trampedach:Stein:2012}. If solar convection actually
would work in an essentially unmixed regime, then which are the critical
Reynolds (and perhaps Prandtl) numbers for the transition away from
the mixing-length regime?  When can we expect to see numerical
simulations of this transition, if it exists?

\section{Future directions}
\label{sec:5}

Reliable seismic inferences on the interior structure of prominent solar 
phenomena such as sunspots, supergranulation, convection and meridional 
circulation would have significant consequences not only for our understanding 
of the way the Sun operates but also for Sun-like stars. Understanding 
magneto-convection and the emergence and sustenance of large-scale field in 
the Sun would provide important constraints for dynamo theory. Thus the 
seismic science of the Sun can play a critical role in advancing astrophysics 
as a whole. 

While global helioseismology, the study of large-scale axisymmetric 
structure in the Sun, has met with great success, %the same cannot be 
local helioseismology, which deals with spatially localized features %said of 
such as sunspots and supergranules and convection in the interior, has a long road ahead. Multi-variable
inverse problems are inevitable in local helioseismology, where seismic 
measurements, obtained at the photosphere, must be related to a large 
number of parameters, e.g., the sound speed, flows and magnetic fields 
in the interior. 
%Inferences obtained using local seismic methods, 
%such as time-distance, ring-diagram analysis and holography have yet to 
%claim the confidence with which global inversions are routinely treated.
%Fundamentally, this may be attributed to the lack of theoretical 
%progress in interpreting seismic measurements. 
In particular, the full accounting of systematical effects, finite wavelengths and inversion non-linearity,
which can significantly influence and bias results,
would greatly improve their trustworthiness and accuracy of local helioseismic inferences. 
The future of a theoretically sound seismology 
holds the promise of revealing important insights such as the 
distribution of large-scale Reynolds stresses, the drivers of 
differential rotation and meridional circulation, and ultimately, 
possibly the source of global magnetism in the Sun itself.
%\bfnote[Criticism of local helioseismology ``a bit too broad'']

Meanwhile, comprehensive simulations of radiative (magneto-)convection
in the near-surface layers of the Sun have achieved an impressive
realism. They compare well with observational results 
\citep[e.g.,][]{nordlund09, Stein:2012}
as well as between different
codes \citep{Beeck2012}.

In the deeper layers, the average properties of such simulations 
are similar to the results of mixing-length models and we can 
expect them to make contact with the anelastic simulations soon as well.
Does that mean that the solar convection zone is in a
well-mixed regime in the sense that the low-entropy downflows
are well mixed into the high-entropy upflows? Or are simulations
in the wrong regime owing to their much too small Reynolds numbers?
The latter view seems to be supported by some results of local
helioseismology, but these need to be confirmed before
definite conclusions can be drawn. If they turn out to be correct,
how can we then reconcile the fact that the near-surface simulations
(which have much too low Reynolds numbers as well) are in such an
excellent agreement with observations, so that even significant
corrections of element abundances of fundamental astrophysical
importance can be derived from them \citep{Asplund2009}.

\bigskip\noindent
We dedicate this paper to our colleague and friend 
Dr. Irene Gonz{\'a}lez Hern{\'a}ndez (1969-2014), 
a pioneer of ring-diagram analysis and far-side imaging,
who passed away on 14 February 2014.

\begin{acknowledgements}
MSM is supported by NASA grants NNH09AK14I (Heliophysics SR\&T) and NNX08AI57G 
(Heliophysics Theory Program). The National Center for Atmospheric Research is 
sponsored by the National Science Foundation. MR acknowledges support from the European Research Council
under the European Union’s Seventh Framework Program
(FP/2007-2013)/ERC Grant Agreement no. 307117.
\end{acknowledgements}

\bibliographystyle{aps-nameyear}      % basic style, author-year citations
%\bibliography{mieschrefs,hanasoge,refs_msch,refs_thompson,refs_mroth}
\bibliography{ms}
\nocite{*}

\end{document}